\documentclass[aps,groupedaddress,showpacs,floatfix]{revtex4-1}
\usepackage{amssymb}
\usepackage{amsmath}
\usepackage{graphicx}
\begin{document}

\title{Free Energy Distribution Function of a Random Ising ferromagnet}

\author{Victor Dotsenko$^{\, a,b}$ and Boris Klumov$^c$}

\affiliation{$^a$LPTMC, Universit\'e Paris VI, 75252 Paris, France}

\affiliation{$^b$L.D.\ Landau Institute for Theoretical Physics,
   119334 Moscow, Russia}

\affiliation{$^c$Joint Institute for High Temperatures, Russian Academy of Sciences, Moscow 125412, Russia}

\date{\today}

\begin{abstract}
We study the free energy distribution function of weakly disordered Ising ferromagnet 
in terms of the $D$-dimensional random temperature Ginzburg-Landau Hamiltonian. 
It is shown that besides the usual Gaussian "body" this distribution function 
exhibits non-Gaussian tails  both in the paramagnetic and in the ferromagnetic phases.
Explicit asymptotic expressions for these tails are derived. 
It is demonstrated that the tails  are strongly asymmetric:
the left tail (for large negative values of the free energy) is much more slow
than the right one (for large positive values of the free energy).
It is argued that in the critical point the free energy of the random Ising ferromagnet
in dimensions $D <4$ is   described by a non-trivial 
universal distribution function being non self-averaging.
\end{abstract}


\maketitle

\medskip

\section{Introduction}

The ferromagnetic systems with weak disorder (e.g. with small concentration of non-magnetic impurities) 
are under intensive study during last four decades. As the disorder of this type
can not  destroy the ferromagnetic ground state,
most of the efforts (both theoretical, experimental and numerical) were concentrated on the 
critical properties of such systems (see e.g. \cite{RandCrit, book}) where even weak disorder
was proved to be "relevant variable". It turns out that the presence of quenched disorder
can essentially modify the critical behavior of the system such that new universal
critical exponents may set in. Another interesting phenomena which appear due to  
quenched randomness in the Ising ferromagnet are the 
so called Griffiths singularities in the thermodynamical functions \cite{Griffiths}.
Apart from these two big issues the effects produced by weak quenched disorder on thermodynamical
properties of a ferromagnet (away from the critical point) were usually considered as 
more or less irrelevant corrections to the pure case.

The aim of the present study is to demonstrate that due to the presence of weak disorder
the statistical properties of
(random) free energy of the Ising ferromagnet 
are getting rather non-trivial. Of course, in the thermodynamic limit 
due to self-averaging the free energy distribution function turns into the 
$\delta$-function. However, if the volume of the system is large but finite
(which, in particular, is the case in numerical simulations)  
the situation is much more complicated. Away from the critical point at scales much bigger than
the correlation length $R_{c}$ the system could be considered as a set of more or less independent 
regions with the size $R_{c}$. In this case one would naively expect that the 
free energy distribution function must be Gaussian. In fact this is not the case.
It can be shown that besides the central Gaussian part this distribution
has asymmetric and essentially non-Gaussian tails. 
    Approaching the critical point  one finds that the range of validity 
of the Gaussian "body"  shrinks while the "tails" are getting of the same order as the "body".
Moreover, if we naively imitate the critical point by assuming that 
the correlation length becomes of the order of the system size 
we formally find that  the free energy distribution 
function turns into a universal curve.

\vspace{5mm}

In our present study the Ising ferromagnet with quenched disorder 
will be considered in terms of the random temperature 
$D$-dimensional Ginzburg-Landau Hamiltonian:
\begin{equation}
 \label{1}
H\bigl[\phi,\xi\bigr] = \int d^{D}{\bf x} \Bigl[\frac{1}{2} \bigl(\nabla \phi({\bf x})\bigr)^{2}
+ \frac{1}{2} (\tau - \xi({\bf x})) \phi^{2}({\bf x})
+ \frac{1}{4} g_{o} \, \phi^{4}({\bf x}) \Bigr]
\end{equation}
where independent random quenched parameters $\xi({\bf x})$ are described by the Gaussian distribution function
\begin{equation}
 \label{2}
P[\xi({\bf x})] \; = \; \prod_{{\bf x}} \Biggl[\frac{1}{\sqrt{4\pi u_{0}}} 
\exp\Bigl(-\frac{1}{4u_{o}} \xi^{2}({\bf x}) \Bigr) \Biggr]
\end{equation}
where $u_{o}$ is the parameter which describes the strength of the disorder,
$\overline{\xi^{2}} = 2 u_{o}$.
In what follows it will be supposed
that the disorder is weak: $u_{o} \ll \tau^{2}$ and $u_{o} \ll g_{o}$.

For a given realization of the disorder the partition function of the considered system is
\begin{equation}\
 \label{3}
Z[\xi] \; = \; \int {\cal D} \phi \; \exp\bigl(-H\bigl[\phi,\xi\bigr]\bigr) \; = \; 
\exp\bigl(-F\bigl[\xi\bigr]\bigr) 
\end{equation}
where $\int {\cal D}\phi$ denotes the integration over all configurations of the fields $\phi({\bf x})$
and $F\bigl[\xi\bigr]$ is the free energy of the system.

The distribution function of the random quantity $F\bigl[\xi\bigr]$ can be analyzed by studying
the moments of the partition function. Taking the integer $n$-th power of the expression in eq.(\ref{3})
and averaging over the disorder parameters $\xi({\bf x})$ (with the Gaussian distribution, eq.(\ref{2}))
we get the replica partition function
\begin{equation}
 \label{4}
\overline{Z^{n}[\xi]} \; \equiv Z(n) \; = 
\int {\cal D} \phi_{1} ... \int {\cal D} \phi_{n}
\exp\bigl(-H_{n}\bigl[\boldsymbol{\phi}\bigr]\bigr)
\end{equation}
where the replica Hamiltonian 
\begin{equation}
 \label{5}
H_{n}\bigl[\boldsymbol{\phi}\bigr] \; = \; 
\int d^{D}{\bf x} \Bigl[ \frac{1}{2} \sum_{a=1}^{n}\bigl(\nabla \phi_{a}({\bf x})\bigr)^{2}
+ \frac{1}{2} \tau \sum_{a=1}^{n} \phi_{a}^{2}({\bf x})
+ \frac{1}{4} \sum_{a,b=1}^{n}  g_{ab} \, \phi_{a}^{2}({\bf x}) \phi_{b}^{2}({\bf x})\Bigr]
\end{equation}
depends on $n$ interacting fields $\boldsymbol{\phi} \equiv \{\phi_{1}, ...\phi_{n}\}$,
and the coupling $(n\times n)$ matrix
\begin{equation}
 \label{6}
g_{ab} \; = \; g_{o} \delta_{ab} \; - \; u_{o}
\end{equation}
Taking into account the definition of the free energy in eq.(\ref{3}), the replica partition  
function, eq.(\ref{4}), can also be represented as follows,
\begin{equation}
 \label{7}
Z(n) \; = \; \overline{\exp\bigl(-n F\bigl[\xi\bigr]\bigr) }
\end{equation}
or
\begin{equation}
 \label{8}
Z(n) \; = \; \int_{-\infty}^{+\infty} dF \, P(F) \, \exp\bigl(-n F\bigr)
\end{equation}
where $P(F)$ is a free energy distribution function.

Usually the replica partition function, eq.(\ref{4}), is studied for deriving 
the  average value of the free energy. The heuristic (not well justified) procedure
of the replica calculations requires to perform the analytic continuation 
of the function $Z(n)$ from integer to arbitrary values of the replica parameter $n$.
Then, taking the limit $n \to 0$ in both sides of eq.(\ref{8}) (and taking 
into account that $\int dF P(F) =1$) we formally get
\begin{equation}
 \label{9}
\lim_{n\to 0} \frac{Z(n) - 1}{n} \; = \;  
\int_{-\infty}^{+\infty} dF  \, P(F) \, F \, \equiv \, \overline{F}
\end{equation}

In fact if we are interested not just in the average free energy of the system but
in the properties of its distribution function the situation even with weak disorder 
turns out to be rather nontrivial. Considering the system away from the critical
point one would naively expect that at least at scales greater than the correlation length
(where the system is expected to split into a set of more or less independent 
"cells" of the size of the correlation length) "everything must be Gaussian distributed".
One can easily prove that this is not true. Indeed, let us consider the coupling matrix
$g_{ab}$, eq(\ref{6}). It has $(n-1)$ eigenvalues $\lambda_{1} = g_{o} - u_{o}$ and 
one eigenvalue $\lambda_{2} = g_{o} - n u_{o}$. We see that 
for $n > g_{o}/u_{o}$ the eigenvalue $\lambda_{2}$ is getting negative. 
Thus, according to eqs.(\ref{4})-(\ref{5}), we find that all moments of the 
partition function $Z(n)$ with $n > g_{o}/u_{o}$ must be divergent! On the other hand,
according to the relation (\ref{8}) this indicate that the free energy distribution function
of the considered system can not be Gaussian. 
Moreover, since it is evident that the divergency in the integration in the rhs of 
eq.(\ref{8}) takes place in the limit $F \to -\infty$, we can easily guess that the 
left asymptotic of the distribution function $P(F)$ has the following simple form:
\begin{equation}
 \label{10}
P(F \to -\infty) \; \sim \; \exp\Bigl(-\frac{g_{o}}{u_{o}} \big| F \big| \Bigr)
\end{equation}
It should be stressed that this phenomenon must be quite general:
it take place for any dimension of the system independently of the value
of the temperature parameter $\tau$ (except for the critical point) both
in paramagnetic and in the ferromagnetic phases.

\section{Toy Model}

To analyze the properties of the free energy distribution function of a random 
ferromagnet one can start with very simple "toy" model. Let us consider 
the system described by the Hamiltonian containing only one degree of freedom:
\begin{equation}
 \label{11}
H(\phi,\xi) =  \frac{1}{2} (\tau - \xi) \phi^{2}
+ \frac{1}{4} g_{o} \, \phi^{4}
\end{equation}
The random quenched parameter $\xi$ is described by the Gaussian distribution function
\begin{equation}
 \label{12}
P(\xi) \; = \; \frac{1}{\sqrt{4\pi u}} 
\exp\Bigl(-\frac{1}{4u_{o}} \xi^{2} \Bigr) 
\end{equation}
and its typical value $\overline{\xi^{2}} = 2 u_{o}$ is supposed to be small,
 $u_{o} \ll \tau^{2}$ and $u_{o} \ll g_{o}$. Besides, to simplify the calculations 
it will be assumed that the  coupling parameter $g_{o}$ is also small,
$g_{o} \ll \tau^{2}$.

\subsection{"Paramagnetic" region ($\tau > 0$)}

To simplify notations let us redefine: 
$\xi \to \tau \xi$, $u_{o} \to \tau^{2} u$, $g_{o} \to \tau^{2} g$ and
$\phi \to \phi/\sqrt{\tau}$. It these new notations the 
partition function of the system  is
\begin{equation}
 \label{13}
Z_{para}(\xi) \; = \; \int_{-\infty}^{+\infty} \frac{d\phi}{\sqrt{2\pi}}
\exp\Bigl[-\frac{1}{2}(1-\xi) \phi^{2} - \frac{1}{4} g \phi^{4} \Bigr] \; = \; 
\exp\bigl[ - F_{para}(\xi)\bigr]
\end{equation}
where $F_{para}(\xi)$ is the (random) free energy of the system. For the replica partition function
we get:
\begin{equation}
 \label{14}
Z_{para}(n) \; = \; \int_{-\infty}^{+\infty} \frac{d\xi}{\sqrt{4\pi u}}
\exp\Bigl(-\frac{1}{4u} \xi^{2} \Bigr) \; \Bigl(Z_{para}(\xi) \Bigr)^{n}
\end{equation}
On the other hand,
\begin{equation}
 \label{15}
Z_{para}(n) \; = \; \int_{-\infty}^{+\infty}  dF \, P_{para}(F) \, \exp\bigl[ - n F\bigr]
\end{equation}
where $P_{para}(F)$ is a free energy distribution function. The above relation is the 
bilateral Laplace transform. Performing analytic continuation of the function
$Z_{para}(n)$ from integer $n$  to arbitrary complex values, by  inverse Laplace
transform we obtain:
\begin{equation}
 \label{16}
P_{para}(F) \; = \; \int_{-i\infty}^{+i\infty}  \frac{dz}{2\pi i}
Z_{para}(z) \,  \exp\bigl[ z\,  F\bigr]
\end{equation}
Under conditions $u \ll g \ll 1$ one can easily perform (approximate)
calculations of the partition function, eq.(\ref{13}). For $\xi < 1$ the 
main contribution in the integration over $\phi$ comes from the vicinity
of the point $\phi = 0$.  On the other hand, at $\xi > 1$ the integral
is dominated by the vicinity of the "ferromagnetic"  points $\phi = \pm \sqrt{(\xi-1)/g}$.
Thus,
\begin{equation}
 \label{17}
Z_{para}(\xi) \; \simeq \; \exp\Bigl[-f_{0}(\xi) + \frac{(\xi - 1)^{2}}{4 g} \theta(\xi -1)\Bigr]
\end{equation}
where
\begin{equation}
 \label{18}
f_{0}(\xi) \; \simeq \;  \left\{ 
                          \begin{array}{ll}
\frac{1}{2} \log|\xi - 1| , \; \; \; \;  \mbox{for} \; \;  |\xi - 1| \gg \sqrt{g} \\
\\
\frac{1}{4}\log(g) , \; \; \; \; \; \; \; \; \; \; \mbox{for} \; \;  |\xi - 1| \ll \sqrt{g}
                        \end{array}
 \right.
\end{equation}
Correspondingly, for the replica partition function, eq.(\ref{14}), we find
\begin{equation}
 \label{19}
Z_{para}(n) \; \simeq \; 
\int_{-\infty}^{+\infty} \frac{d\xi}{\sqrt{4\pi u}}
\exp\Bigl[-\frac{1}{4u} \xi^{2}  - n F_{para}(\xi) \Bigr]
\end{equation}
where
\begin{equation}
 \label{20}
F_{para}(\xi) \; \simeq \; f_{0}(\xi) - \frac{(\xi - 1)^{2}}{4 g} \theta(\xi -1)
\end{equation}
One can easily note here that since 
$F_{para}(\xi \to +\infty) \sim - \frac{1}{4 g} \xi^{2}$
the partition function $Z_{para}(n)$ in eq.(\ref{19}) is divergent for all values
$ n > g/u$. 

Finally, substituting eqs.(\ref{19})-(\ref{20}) into eq.(\ref{16}) (where the integration contour 
can be chosen to be just the imaginary axes, $z = i\omega$) we obtain
\begin{eqnarray}
 \nonumber
P_{para}(F) &=& 
\int_{-\infty}^{+\infty}  \frac{d\omega}{2\pi} \,
\int_{-\infty}^{+\infty} \frac{d\xi}{\sqrt{4\pi u}}
\exp\Bigl[-\frac{1}{4u} \xi^{2} + i\omega \bigl(F - F_{para}(\xi)\bigr) \Bigr]
\\
\nonumber
\\
&=& 
\int_{-\infty}^{+\infty} \frac{d\xi}{\sqrt{4\pi u}}
\exp\Bigl[-\frac{1}{4u} \xi^{2}\Bigr] \; \delta\Bigl(F - F_{para}(\xi)\Bigr)
\label{21}
\end{eqnarray}

Right part of this distribution function $P_{para}^{(+)}(F)$ valid for 
$-\frac{1}{4}\log(1/g) \ll F < +\infty$ is defined by the region
$-\infty < (\xi -1) \ll - \sqrt{g}$, where $F_{para}(\xi) \simeq \frac{1}{2} \log(1-\xi)$.
Using eq.(\ref{21}) we get 
\begin{equation}
\label{22}
P_{para}^{(+)}(F) \; \simeq \; \frac{1}{\sqrt{\pi u}} 
\exp\Bigl[ 2F -\frac{1}{4u} \Bigl(\exp\bigl(2 F\bigr) - 1\Bigr)^{2} \Bigr]
\end{equation}
In the vicinity of the point $F_{*} = u$ the above distribution has the Gaussian
form:
\begin{equation}
 \label{23}
P_{para}^{(G)}(F) \; \simeq \; \frac{1}{\sqrt{\pi u}} 
\exp\Bigl[-\frac{1}{u} \bigl(F - u\bigr)^{2} \Bigr]
\end{equation}
However, the above form of the distribution is valid only for 
$|F| \ll 1$. For large (positive) values of the free energy
the distribution becomes essentially non-Gaussian:
\begin{equation}
 \label{24}
P_{para}(F \to +\infty) \; \sim \; 
\exp\Bigl[ -\frac{1}{4u} \exp\bigl(4 F\bigr) \Bigr]
\end{equation}
which goes to zero much faster than the Gaussian curve.

Left asymptotic of the distribution function,  $P_{para}^{(-)}(F)$
valid for 
$-\infty \ll F \ll -\frac{1}{4}\log(1/g) $ is defined by the region
$\sqrt{g} \ll (\xi -1) < +\infty$, where $F_{para}(\xi) \simeq -\frac{1}{4g} (\xi-1)^{2}$.
Using eq.(\ref{21}) we get:
\begin{equation}
 \label{25}
P_{para}^{(-)}(F) \; \simeq \; 
\sqrt{\frac{g}{4\pi u |F|}} \, 
\exp\Bigl[- \frac{1}{4u}\bigl(1 + \sqrt{4 g |F|}\bigr)^{2} \Bigr]
\end{equation}
Thus we find that the left tail of the free energy distribution function
$P_{para}(F)$ is also non-Gaussian: 
\begin{equation}
 \label{26}
P_{para}(F\to -\infty) \; \sim \; \exp\Bigl( -\frac{g}{u} |F|\Bigr)
\end{equation}
and it goes to zero much slower that the Gaussian curve. 

In conclusion we find that the free energy distribution function
of the "toy" model, eq.(\ref{11}), at $\tau >0$ is essentially non-Gaussian
and strongly asymmetric: its right tail, eq.(\ref{24}), is much steeper
than the left one, eq.(\ref{26}). 
The example of the entire distribution function
$P_{para}(F)$ for $g =0.1$ and $u = 0.01$ is shown in Figure 1.

\begin{figure}[h]
\begin{center}
   \includegraphics[width=12.0cm]{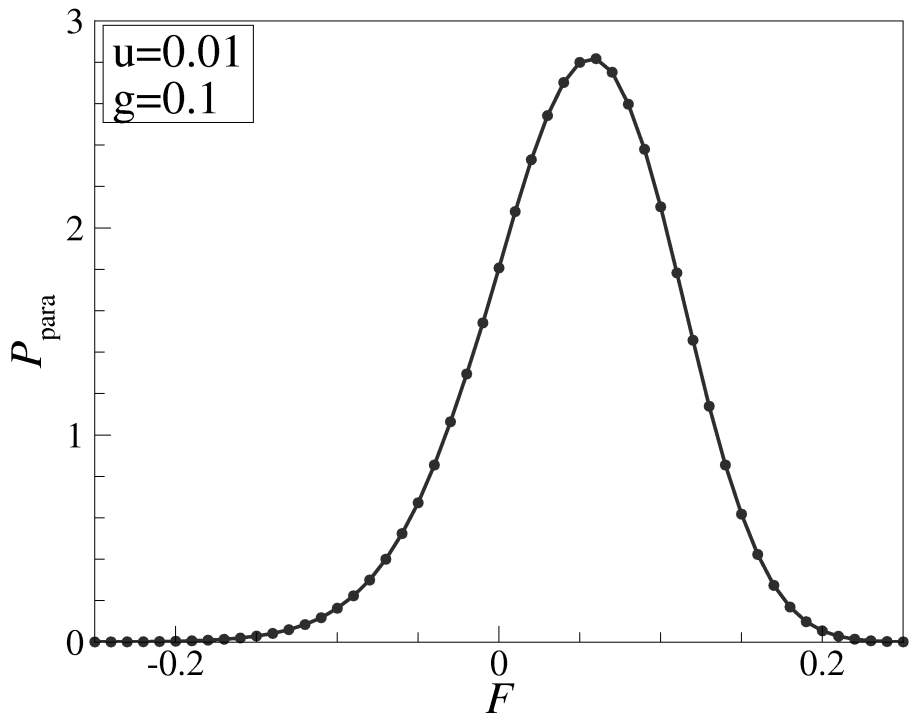}
\caption[]{Free energy distribution function of the "paramagnetic"
toy model, eq.(\ref{13}), with $g=0.1$ and $u=0.01$}
\end{center}
\label{figure1}
\end{figure}

\subsection{"Ferromagnetic" region ($\tau < 0$)}

For the negative values of the temperature parameter $\tau$ in the toy model,
eq.(\ref{11}), we redefine: 
$\xi \to -|\tau| \xi$, $u_{o} \to \tau^{2} u$, $g_{o} \to \tau^{2} g$ and
$\phi \to \phi/\sqrt{|\tau|}$. It these new notations the 
partition function of the system becomes
\begin{equation}
 \label{27}
Z_{ferro}(\xi) \; = \; \int_{-\infty}^{+\infty} \frac{d\phi}{\sqrt{2\pi}}
\exp\Bigl[\, \frac{1}{2}(1 - \xi) \phi^{2} - \frac{1}{4} g \phi^{4} \Bigr] \; = \; 
\exp\bigl[ - F_{ferro}(\xi)\bigr]
\end{equation}
where random parameters $\xi$ are described by the Gaussian distribution,
eq.(\ref{12}). The calculations for the "ferromagnetic" case 
are quite similar to the case $\tau > 0$,
considered in the previous subsection. In particular, for the partition function we get:
\begin{equation}
 \label{28}
Z_{ferro}(n) \; \simeq \; 
\int_{-\infty}^{+\infty} \frac{d\xi}{\sqrt{4\pi u}}
\exp\Bigl[-\frac{1}{4u} \xi^{2}  - n F_{ferro}(\xi) \Bigr]
\end{equation}
where
\begin{equation}
 \label{29}
F_{ferro}(\xi) \; \simeq \; f_{0}(\xi) - \frac{(1-\xi)^{2}}{4 g} \theta(1-\xi)
\end{equation}
Here $f_{0}(\xi) $ is defined in eq.(\ref{18}). As in  the "paramagnetic"
case, one can easily note that since 
$F_{ferro}(\xi \to -\infty) \sim -\xi^{2}/4g$, 
the partition function $Z_{ferro}(n)$, eq.(\ref{28}), with $n > g/u$ is divergent. 

Similarly to eq.(\ref{21}), for the free energy distribution function 
we get:

\begin{eqnarray}
 \nonumber
P_{ferro}(F) &=& 
\int_{-\infty}^{+\infty}  \frac{d\omega}{2\pi} \,
\int_{-\infty}^{+\infty} \frac{d\xi}{\sqrt{4\pi u}}
\exp\Bigl[-\frac{1}{4u} \xi^{2} + i\omega \bigl(F - F_{ferro}(\xi)\bigr) \Bigr]
\\
\nonumber
\\
&=& 
\int_{-\infty}^{+\infty} \frac{d\xi}{\sqrt{4\pi u}}
\exp\Bigl[-\frac{1}{4u} \xi^{2}\Bigr] \; \delta\Bigl(F - F_{ferro}(\xi)\Bigr)
\label{30}
\end{eqnarray}
The right tail of this function, $P_{ferro}^{(+)}(F)$, valid for
$-\frac{1}{4}\log(1/g) \ll F < +\infty$ is defined by the region $\sqrt{g} \ll (\xi - 1) < +\infty$
where $F_{ferro}(\xi) \simeq \frac{1}{2}\log(\xi - 1)$.
Using eq.(\ref{30}) we find

\begin{equation}
\label{31}
P_{ferro}^{(+)}(F) \; \simeq \; \frac{1}{\sqrt{\pi u}} 
\exp\Bigl[ 2F -\frac{1}{4u} \Bigl(\exp\bigl(2 F\bigr) + 1\Bigr)^{2} \Bigr]
\end{equation}
Thus the decay of the right tail of this function coincides with 
the one of the "paramagnetic" function, eq(\ref{24}),
\begin{equation}
 \label{32}
P_{ferro}(F \to +\infty) \; \sim \; 
\exp\Bigl[ -\frac{1}{4u} \exp\bigl(4 F\bigr) \Bigr]
\end{equation}
The left tail of the free energy distribution function, $P_{ferro}^{(-)}(F)$, valid for
$-\infty < F \ll -\frac{1}{4}\log(1/g)$ is defined by the region $-\infty < (\xi - 1) \ll -\sqrt{g}$
 where $F_{ferro}(\xi) \simeq -\frac{1}{4g}(1-\xi)^{2}$.
Using eq.(\ref{30}) we find
\begin{equation}
 \label{33}
P_{ferro}^{(-)}(F) \; \simeq \; 
\sqrt{\frac{g}{4\pi u |F|}} \, 
\exp\Bigl[- \frac{1}{4u}\bigl(1 - \sqrt{4 g |F|}\bigr)^{2} \Bigr]
\end{equation}
In the vicinity of the "ferromagnetic" average free energy $F_{*} = -1/4g$\
this distribution is Gaussian,
\begin{equation}
 \label{34}
P_{ferro}^{(G)}(F) \; \simeq \;
\frac{g}{\sqrt{\pi u}} \exp\Bigl[-\frac{g^{2}}{u} \bigl(F + \frac{1}{4g}\bigr)^{2}\Bigr]
\end{equation}
However at large negative values of the free energy (similarly to the "paramagnetic" case)
the distribution becomes essentially non-Gaussian:
\begin{equation}
 \label{35}
P_{ferro}(F\to -\infty) \; \sim \; \exp\Bigl( -\frac{g}{u} |F|\Bigr)
\end{equation}
The example of the entire distribution function
$P_{ferro}(F)$ for $g =0.1$ and $u = 0.01$ is shown in Figure 2.
Again, like in the "paramagnetic" case this function is strongly asymmetric:
its left tail, eq.(\ref{35}), is much slower that the right one, eq.(\ref{32}).

\begin{figure}[h]
\begin{center}
   \includegraphics[width=12.0cm]{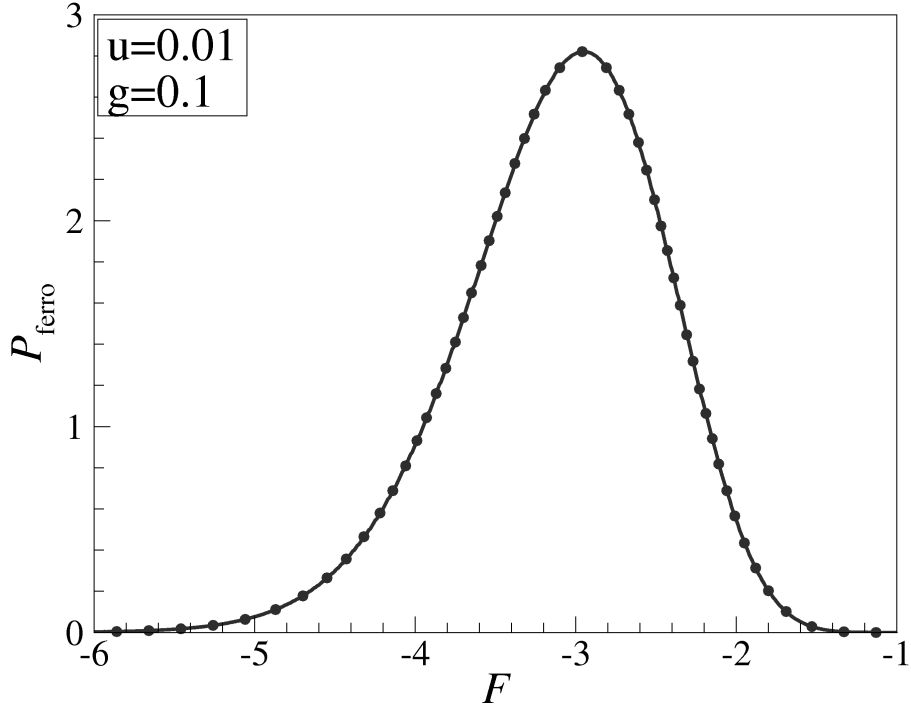}
\caption[]{Free energy distribution function of the "ferromagnetic"
toy model, eq.(\ref{27}), with $g=0.1$ and $u=0.01$}
\end{center}
\label{figure2}
\end{figure}

\section{Macroscopic system}

Let us consider $D$-dimensional random ferromagnet described by the Ginsburg-Landau Hamiltonian,
eq.(\ref{1}). In the mean-field approximation away from the critical point (in the paramagnetic phase,
$\tau > 0$) this system can be considered as a set of $N \simeq V/R_{c}^{D}(\tau)$ independent
"sells", where $V$ is the volume of the system and $R_{c}(\tau) \sim 1/\sqrt{\tau}$ 
is the correlation length. Performing spatial rescaling $x \to R_{c} x = \tau^{-D/2} x$,
 according to the standard ideas of the scaling theory we have to rescale
$\phi \to \tau^{(D-2)/2} \phi$, $g_{o} \to \tau^{-(D-4)/4} g$ and $u_{o} \to \tau^{-(D-4)/2} u$.
Then in terms of the new (rescaled) parameters the partition function of the 
considered system can be estimated as follows
\begin{equation}
 \label{36}
{\cal Z}\bigl[\boldsymbol{\xi}\bigr] \; \simeq \;  
\prod_{k=1}^{N}
\Biggl[\int_{-\infty}^{+\infty} \frac{d\phi}{\sqrt{2\pi}}
\exp\Bigl[-\frac{1}{2}(1-\xi_{k}) \phi^{2} - \frac{1}{4} g \phi^{4} \Bigr] \Biggr] \; = \;
 \prod_{k=1}^{N} Z_{para}(\xi_{k})
\end{equation}
where the function ${\cal Z}\bigl[\boldsymbol{\xi}\bigr]$ depends on $N$ random parameters 
$\boldsymbol{\xi} \equiv \{\xi_{1}, ...\xi_{N}\}$ and 
$Z_{para}(\xi)$ 
is the partition function of the toy model, eq.(\ref{13}). The parameters 
$\{\xi_{k}\}$ are described by the independent 
Gaussian distributions
\begin{equation}
 \label{37}
P[\boldsymbol{\xi}] \; = \; \prod_{k=1}^{N} \Biggl[\frac{1}{\sqrt{4\pi u}} 
\exp\Bigl(-\frac{1}{4u} \xi_{k}^{2} \Bigr) \Biggr]
\end{equation}
Correspondingly, for the replica partition function we obtain
\begin{equation}
 \label{38}
\overline{{\cal Z}^{n}} \equiv {\cal Z}(n) \; \simeq \; 
\Biggl[
\int_{-\infty}^{+\infty} \frac{d\xi}{\sqrt{4\pi u}}
\exp\Bigl(-\frac{1}{4u} \xi_{k}^{2} \Bigr) \; \Bigl(Z_{para}(\xi)\Bigr)^{n} \Biggr]^{N} \; = \; 
\Bigl(Z_{para}(n)\Bigr)^{N}
\end{equation}
where $Z_{para}(n)$ is the replica partition function of the toy model, eq.(\ref{14}).
On the other hand, 
\begin{equation}
 \label{39}
{\cal Z}(n) \; = \; \int_{-\infty}^{+\infty} dF \, {\cal P}(F) \, \exp\bigl(-n F\bigr) 
\end{equation}
where ${\cal P}(F)$ is the total free energy distribution function 
of the considered macroscopic system.
Performing inverse Laplace transform we get
\begin{equation}
 \label{40}
{\cal P}(F) \; = \; \int_{-i\infty}^{+i\infty} \frac{dz}{2\pi i} 
\Bigl(Z_{para}(z)\Bigr)^{N} \exp\bigl(z F\bigr) 
\end{equation}
%t
Taking $z = i\omega$, introducing  the free energy density, $f = F/N$,
and substituting here eq.(\ref{19}),  for the corresponding probability distribution function
we find the following expression
\begin{equation}
 \label{41}
\tilde{{\cal P}}(f) \; = \; \int_{-\infty}^{+\infty} \frac{d\omega}{2\pi} \; 
\Biggl[
\int_{-\infty}^{+\infty} \frac{d\xi}{\sqrt{4\pi u}}
\exp\Bigl(-\frac{1}{4u} \xi^{2} + i\omega \bigl[f - F_{para}(\xi)\bigr]\Bigr)\Biggr]^{N}
\end{equation}
where the function $F_{para}(\xi)$ is defined in eqs.(\ref{20}) and (\ref{18}).

The main Gaussian part of the above distribution valid for the values $|f| \ll 1$
is defined by the region $|\xi| \ll 1$ where
$F_{para}(\xi) \simeq \frac{1}{2} \log(1-\xi) \simeq -\frac{1}{2} \xi$.
In this case
\begin{eqnarray}
\nonumber
\tilde{{\cal P}}^{(G)}(f) & \simeq &
\int_{-\infty}^{+\infty} \frac{d\omega}{2\pi} \;
\Biggl[
\int_{-\infty}^{+\infty} \frac{d\xi}{\sqrt{4\pi u}}
\exp\Bigl(-\frac{1}{4u} \xi^{2} + \frac{i}{2}\omega \xi +i\omega f \Bigr)\Biggr]^{N}
\\
\nonumber
\\
&=& \frac{1}{\sqrt{\pi N u}} \exp\Bigl(-\frac{N}{u} f^{2} \Bigr)
\label{42}
\end{eqnarray}
Correspondingly, for the distribution function of the total free energy $F = f N = f V/R_{c}^{D}$
we get:
\begin{equation}
 \label{43}
{\cal P}^{(G)}(F) \; = \; \sqrt{\frac{R_{c}^{D}}{V \pi u}} \; \exp\Bigl(-\frac{R_{c}^{D}}{V u} F^{2} \Bigr)
\end{equation}
One can easily note that in the thermodynamic limit $V \to \infty$ this distribution function 
turns into the $\delta$-function which assumes that the considered random system is self-averaging.
However, at large but finite volume of the system one can easily prove that both the 
left and the right tails of the function ${\cal P}(F)$ are essentially non-Gaussian.

Right tail $\tilde{{\cal P}}^{(+)}(f)$ of the distribution function, eq.(\ref{41}), valid for the values
$f \gg 1$ is defined by the region $ |\xi| \gg 1$ ($\xi<0$) where
$F_{para}(\xi) \simeq \frac{1}{2} \log|\xi|$. In this case
\begin{eqnarray}
\nonumber
\tilde{{\cal P}}^{(+)}(f) & \simeq &
\prod_{k=1}^{N}\Biggl[\int_{-\infty}^{+\infty} \frac{d\xi_{k}}{\sqrt{4\pi u}} \Biggr] \;
\exp\Bigl[ -\frac{1}{4u} \sum_{k=1}^{N} \xi_{k}^{2} \Bigr] \; 
\int_{-\infty}^{+\infty} \frac{d\omega}{2\pi} \;
\exp\Bigl[ i\omega \sum_{k=1}^{N} \bigl(f - \frac{1}{2}\log|\xi_{k}|\bigr)\Bigr] 
\\
\nonumber
\\
&\sim&
\exp\Bigl[-\frac{N}{4u} \exp\bigl(4 f\bigr)\Bigr] \; = \; 
\exp\Bigl[-\frac{V}{4u R_{c}^{D}} \exp\bigl(4  f \bigr)\Bigr]
\label{44}
\end{eqnarray}
Correspondingly, for the right tail of the total free energy distribution function
we get 
\begin{equation}
 \label{45}
{\cal P}^{(+)}(F) \sim 
\exp\Bigl[-\frac{V}{4u R_{c}^{D}} \exp\Bigl(4 \frac{R_{c}^{D}}{V} F \Bigr)\Bigr]
\end{equation}
which is valid for $F \gg V/R_{c}^{D}$.

Left tail $\tilde{{\cal P}}^{(-)}(f)$ of the distribution function, eq.(\ref{41}), valid for $f < 0$ and
$|f| \gg 1$ is defined by the region $\xi \gg 1$ where
$F_{para}(\xi) \simeq -\frac{1}{4g} \xi^{2}$. In this case
\begin{eqnarray}
\nonumber
\tilde{{\cal P}}^{(-)}(f) & \simeq &
\int_{-i\infty}^{+i\infty} \frac{dz}{2\pi i}
\Biggl[\int_{-\infty}^{+\infty} \frac{d\xi}{\sqrt{4\pi u}} 
\exp\Bigl[ -\frac{1}{4u}\Bigl(1 - \frac{u}{g} z \Bigr) \xi^{2}
\Biggr]^{N} \exp\bigl(N f z\bigr) \;
\\
\nonumber
\\
&=&
\int_{-i\infty}^{+i\infty} \frac{dz}{2\pi i}
\Bigl(1 - \frac{u}{g} z\Bigr)^{-N/2} \; \exp\bigl(- N |f| z\bigr) 
\label{46}
\end{eqnarray}
Shifting here the integration contour to the right by the value $g/u$: $z = g/u + i\omega$
and neglecting pre-exponential factor, we obtain
\begin{equation}
 \label{47}
\tilde{{\cal P}}^{(-)}(f) \; \sim \; 
\exp\Bigl(- N\frac{g}{u} |f|\Bigr) 
\end{equation}
or, for the total free energy (the result already discussed in the Introduction, eq.(\ref{10})), 
\begin{equation}
 \label{48}
{\cal P}^{(-)}(F) \; \sim \; \exp\Bigl(- \frac{g}{u} |F|\Bigr)
\end{equation}
which is valid for $|F| \gg V/R_{c}^{D}$.

Thus, according to eqs.(\ref{43}), (\ref{45}) and (\ref{48}) for the entire
free energy distribution function of the random Ising ferromagnet (in the paramagnetic phase
away from the critical point) we find the following result:
\begin{equation}
 \label{49}
{\cal P}(F) \; \sim \; \left\{ 
                          \begin{array}{ll}
\exp\Bigl(- \frac{g}{u} |F|\Bigr) ; 
\; \; \; \; \; \; \; \; \; \; \; \; \; \; \; \; \; \; \; \; \; \; \; \, 
\mbox{for} \; F < 0 \; \mbox{and} \; |F| \gg \frac{V}{R_{c}^D} 
\\
\\                          
\exp\Bigl(-\frac{R_{c}^{D}}{V u} F^{2} \Bigr) ; 
\; \; \; \; \; \; \; \; \; \; \; \; \; \; \; \; \; \; \; \; \; 
\mbox{for} \;  |F| \ll \frac{V}{R_{c}^D}  
\\
\\
\exp\Bigl[-\frac{V}{4u R_{c}^{D}} \exp\Bigl(4 \frac{R_{c}^{D}}{V} F \Bigr)\Bigr]  ;
\; \; \; \mbox{for} \;  F \gg \frac{V}{R_{c}^D}                    
\end{array}
\right.
\end{equation}
where (in the framework of the mean-field scaling theory)
$R_{c} = 1/\sqrt{\tau}$ is the correlation length, $V$ is the volume of the system, 
$u = \tau^{(D-4)/2} u_{0}$ is the 
rescaled strength of the disorder, eq.(\ref{2}), and $g = \tau^{(D-4)/2} g_{0}$ is the 
rescaled coupling constant of the original Ginsburg-Landau Hamiltonian, eq.(\ref{1}).

One can easily repeat similar calculations for the free energy distribution function
in the ferromagnetic phase to obtain the result similar to eq.(\ref{49}). 
The only difference is that in this case the Gaussian central part of the distribution
will be concentrated around the ferromagnetic ground state energy 
$F_{0} = -\frac{1}{4g} \big(V/R_{c}^{D}) \; = \; -\frac{\tau^{2}}{4g_{0}} V$.

\section{Conclusions}

One can easily note at least three apparent consequences of the obtained result,
eq.(\ref{49}), for the free energy distribution function of the  random
Ising ferromagnet.

First. In the thermodynamic limit, $V \to \infty$, this distribution turns into
the $\delta$-function. This is not surprising and it just indicates that the considered system
is self-averaging.

Second. At large but finite volume of the system in addition to the trivial Gaussian "body"
the distribution function exhibits essentially non-Gaussian tails. Moreover,
these tails are strongly asymmetric: the left tail, at $F \to -\infty$, is much slower 
that the right one, at $F \to +\infty$.  
It is this type of asymmetry of the free energy distribution function 
(the left tail is slow while the right tail is fast) 
which is observed in others random systems \cite{UnivRand, replicas, GausRandForce, DirPolyTW}
It is interesting to note that the result, eq.(\ref{49}), has to be 
very general taking place at all temperatures (except for the critical region)
and in all dimensions. On the other hand the behavior of each tail (both left and right)
looks rather universal. Indeed, at large negative values of the free energy
the logarithm of the distribution function is expected to be {\it linear} in $|F|$,
while at large positive energies, it is the {\it double logarithm} of the distribution function
which is expected to be linear in $F$. Presumably these two types of asymptotics could be 
relatively easy recovered by computer simulations.

Third. While approaching the critical point  we observe that the range of validity 
of the Gaussian "body" of the 
distribution function shrinks while the "tails" are getting of the same order as the "body".
Of course, simple mean-field calculations considered in this paper are not valid in the critical point.
Nevertheless, if we naively imitate the critical point by assuming that 
the correlation length becomes of the order of the system size (i.e. $R_{c}^{D} \sim V$)
while the coupling parameters $g$ and $u$ (in dimension $D < 4$)  become equal to
their universal (depending only of the dimension $D$) fixed point values $g_{*}$ and $u_{*}$, 
\cite{RandCrit} we find that (in the thermodynamic limit $V\to\infty$) the free energy distribution 
function turns into the {\it universal} curve such that
\begin{equation}
 \label{50}
{\cal P}_{crit}(F) \; \sim \; \left\{ 
                          \begin{array}{ll}
\exp\Bigl(- \frac{g_{*}}{u_{*}} |F|\Bigr) ; 
\, \; \; \; \; \; \; \; \; \; \; \; \; \; \,
\mbox{for} \; F < 0 \; \mbox{and} \; |F| \gg 1
\\
\\                          
\exp\Bigl(-\frac{1}{u_{*}} F^{2} \Bigr) ; 
\, \; \; \; \; \; \; \; \; \; \; \; \; \; \; 
\mbox{for} \;  |F| \ll 1  
\\
\\
\exp\Bigl[-\frac{1}{4u_{*}} \exp\Bigl(4  F \Bigr)\Bigr]  ;
\; \; \; \mbox{for} \;  F \gg 1                    
\end{array}
\right.
\end{equation}
All that evokes the old idea that free energy (as well as some others thermodynamic
quantities) of the random ferromagnets could be non self-averaging in the critical
point \cite{NonSelfAverage}. Of course, the systematic derivation of the free energy distribution function 
in the critical point requires a special consideration.

\acknowledgments

This work was supported in part by the grant IRSES DCPA PhysBio-269139

\end{document}